# LEXICAL ANALYSIS OF AUTOMATED ACCOUNTS ON TWITTER

Isa Inuwa-Dutse[1], Bello Shehu Bello[2] and Ioannis Korkontzelos[1]
[1]Department of Computer Science, Edge Hill University, UK
[2]Department of Informatics, University of Leicester, UK

**ABSTRACT**

In recent years, social bots have been using increasingly more sophisticated, challenging detection strategies. While many approaches and features have been proposed, social bots evade detection and interact much like humans making it difficult to distinguish real human accounts from bot accounts. For detection systems, various features under the broader categories of account profile, tweet content, network and temporal pattern have been utilised. The use of tweet content features is limited to analysis of basic terms such as URLs, hashtags, name entities and sentiment. Given a set of tweet contents with no obvious pattern can we distinguish contents produced by social bots from that of humans? We aim to answer this question by analysing the lexical richness of tweets produced by the respective accounts using large collections of different datasets. Our results show a clear margin between the two classes in lexical diversity, lexical sophistication and distribution of emoticons. We found that the proposed lexical features significantly improve the performance of classifying both account types. These features are useful for training a standard machine learning classifier for effective detection of social bot accounts. A new dataset is made freely available for further exploration.

**KEYWORDS**

Twitter Bots, Lexical Analysis, Bot Detection, Social Network Analysis

## 1. INTRODUCTION

As a social being, human's behaviour is largely influenced by close associates. In this era of modern civilisation, the Internet has been the precursor of the socialisation being witnessed today. Social networking sites are important avenues for instant information sharing and interaction on a large scale (Gilani et al., 2017; Wilson et al., 2012). Modern social media platforms, such as Twitter and Facebook, enable various forms of interactions among diverse users. This capability results in a huge amount of data waiting to be mined by researchers. However, the credibility of such content is being questioned by the growing activities of automated accounts otherwise known as social bots. *Social Bots* are automated accounts designed to follow a well-defined algorithm to interact with users either by amplifying or generating contents in social media platforms (Ferrara et al., 2016). Some social bot accounts are legitimate, with clear distinguishing features, such as *@congressedits*[1], whereas the majority are created to mislead in various ways, such as by creating superficial popularity (Varol et al., 2017) or influencing public opinion (Howard and Kollanyi, 2016). Some bots are obvious, because they use the word "bot" explicitly in their Twitter handle (Inuwa-Dutse et al., 2018). Autonomous accounts contribute a sizeable part of social media content. It was estimated that 9% - 15% of active Twitter accounts are social bots accounts (Varol et al., 2017) and require effective methods to be detected.

How could we detect if a given tweet was posted by a social bot or a human user? We investigate this question based on a collection of tweets sampled from Twitter. In particular, we investigate how effective lexical features are for the detection of social bot accounts. In contrast to previous work (Benevenuto et al., 2010; Cai t al., 2017; Dockerson et al., 2014; Lee and Kim, 2012; Thomas et al., 2011; Varol et al., 2017; Wang, 2010), our study focuses on comprehensive linguistic analysis to define lexical features effective enough for accurate detection of social bot accounts on Twitter. As shown in Table 1, the study explores a large number of tweets from social bots and humans to understand the difference between the two in terms of lexical richness and distribution of emoticons, further discussed in section 3. Our analysis highlights the distinguishing characteristics of automated accounts and how lexical features can improve detection of bots.

---

[1] A bot that tweets anonymous Wikipedia edits made from IP addresses in the US Congress.

*Contribution:* The study contributes a powerful set of features useful in distinguishing between humans and social bots on Twitter. We provide the first comprehensive analysis of lexical richness of tweets computed on various accounts and investigate how their incorporation improves the performance of the detection system. Features based on *lexical diversity, type-token ratio, usage of contractions and emoticons* are powerful lexical signals in distinguishing between humans and social bots. The study provides a new set of distinctive features and a dataset to support the research community in identifying bot accounts.

The remaining of the paper is structured as follows. Section 2 and Section 3 present a review of related works and propose lexical features, respectively. Section 4 presents our experiments and Section 5 presents the results and a detailed discussion. Finally, Section 6 concludes the study and proposes some future work.

## 2. RELATED WORK

Social bot accounts are instrumental in spreading fake and malicious news on Twitter, employed to skew analysis results and opinion of users. The demand for effective detection systems has prompted a surge of various research approaches. Early social bot accounts have been reported to lack basic account information such as meaningful screen names or profile picture (Varol et al., 2017; Lee et al., 2011). This is no longer the case, as social bots grow in sophistication, making it difficult to identify distinguishing features from human accounts. Some approaches involve the analysis of accounts as far as their position in a *network*, their *temporal* metadata and the *content of their tweets*, to define a new set of features. The following review focuses on related studies that utilised aspects of these features to detect bot accounts.

*Network and Temporal Features:* The study of Chavoshi et al. (2017) analyses the behavioural patterns of accounts by focusing on features related to the network structure, such as local motifs, i.e. repeating behaviour, and discords, i.e. anomalous behaviour. The study also shows how temporal behaviour is useful as a means to distinguish bot from human user accounts. The work of Davis et al. (2016) and (Ferrara et al. (2016) developed a detection system that leverages features related to both network structure and tweeting behaviour exhibited by accounts. In [**Error! Reference source not found.**] researchers utilised many features related to network, temporal behaviour, tweet syntax, tweet semantics and user profile for the detection of *influence bots*. Influence bots are categories of social bot accounts that aim at influencing the opinion of other users. However, despite the wide spectrum of features considered in the study, analysis of the lexical richness was not covered.

*Tweet Content:* Many detection systems have been developed by leveraging the content of tweets posted by account holders on Twitter. This approach was adopted in Dockerson et al. (2014) to detect social bot accounts on Twitter based on sentiment features, such as topic sentiment and the transition frequency of tweet's sentiment, to train a machine learning classifier. Similarly, to the *tweet content* approach that relies on linguistic analysis, Inuwa-Dutse et al. (2018) utilised features based on lexical richness to detect spam accounts on Twitter. The study of (Cai t al., 2017) proposed a deep learning approach that incorporates content and behavioural information to detect social bots.

Related literature pays little attention to the analysis of lexical richness of various users' tweets and how they can inform the detection of social bots on Twitter. In contrast to previous work, this study is based on in-depth analysis of lexical richness as a basis for building a detection system. We present an approach solely based on lexical analysis of contents from both humans and social bots accounts to distinguish between them.

## 3. LEXICAL RICHNESS FEATURES

This section describes the lexical features utilised in the study, able to improve detection systems. Lexical richness is a broad concept, expressed in various forms and metrics to assess the quality of text. Metrics such as lexical diversity, lexical sophistication and lexical density are commonly used in linguistic analysis (Šišková, 2012; Templin, 1957). Our approach is based on lexical richness of tweets from various Twitter accounts to detect social bots.

## 3.1 Type-Token Ratio

Type-Token Ratio (TTR) is a simple, yet powerful metric used commonly in quantitative linguistics to measure the richness of vocabulary in a given context (Tweedie and Baayen, 1998). TTR can be expressed as $V(N)/N$, i.e. the size of vocabulary in $N$ divided by the total size of N. In the context of this study, TTR is the ratio of unique tokens in a tweet to the total number of tokens in the tweet. It can be argued that computing TTR over all tweets of an account may lead to a better result. However, the small size of individual tweets may skew the result due to the sparsity of unique tokens relative to the total number of tokens.

## 3.2 Lexical Diversity

Lexical Diversity is an important metric in the analysis of lexical richness. It is useful in assessing the distribution of different content words[2] utilised in a textual corpus or in speech (Tweedie and Baayen, 1998). Lexical sophistication is similar to lexical diversity and focuses on understanding the distribution of advanced words. This study focuses on lexical diversity. The rationale behind using it as a feature is to assess its levels in bot and human accounts and its predictive power in detection of social bot accounts. While the contents produced by social bots were shown to be different across various social bots (Morstatter et al., 2016), they tend to exhibit similarities in terms of widespread use of URLs. In view of this, we computed lexical diversity as the total number of tokens in a tweet without URLs, user mentions and stopwords divided by the total number of tokens in the tweet.

## 3.3 Usage of Contractions

Text or speech in English can be shortened by ignoring some letters or phonetics. These kinds of contracted words are a form of lexical sophistication, useful to measure the fluency of users. Various forms of contracted words are widespread on Twitter. However, we focus on a predefined list of contractions[3] for our analysis. The expectation is for human users to use diverse contractions, while it would be difficult for a bot to use contraction unless generating its tweet from pre-existing sources, e.g. a book or a structured document.

## 3.4 Emoticons

Emoticons[4] are collections of pictorial representations of facial expressions or *emojis* in form of various characters (letters, punctuation, and numbers) to convey emotional mood. Emoticons are popular on social media, especially on Twitter, where tweets are of limited length, are useful indicators or users' sentiment. Common examples of *emoticons* are the smiley, *:-)*, and the sad face, *:-(*. Sentiment-related features have been shown to contribute in detection systems (Dockerson et al., 2014). We leverage this to understand the role of *emoticons* in detection of bot accounts using a comprehensive list of emoticons[5]. The goal is to understand how human and social bot users apply emoticons in tweets and utilise the insight to build classification models. We hypothesise human users will use emoticons in a larger proportion in comparison to social bots accounts.

## 4. EXPERIMENT

This section describes the datasets utilised in this study, including the collection procedure and pre-processing. This is followed by feature selection and building the classification framework.

## 4.1 Dataset

---
[2] Words with meaning in a text; not in stopwords.
[3] Wikipedia list of contractions: en.wikipedia.org/wiki/Wikipedia:List_of_English_contractions
[4] A portmanteau of *emotion* and *icon*.
[5] Available from: en.wikipedia.org/wiki/Emoticon

We utilized three different datasets. The first two are publicly available, *Dataset1* is obtained from (Morstatter et al., 2016) and *Dataset2* from (Gilani et al., 2017. The last dataset is collected for the purpose of this study. Table 1 summarises the datasets. *Dataset2* is classified under five different groups based on popularity and volume of contents generated by the accounts. We maintain the groupings in the study and analyse the lexical richness in each group to understand how the lexical richness will vary across the different groups. The two datasets contain some non-English accounts, which were removed to facilitate proper lexical analysis of tweets. *Dataset1* provides only account *ids* and *Dataset2* provides screen-names and account features. We used the Twitter API to crawl tweets from each account. Our analysis experience with *Dataset1* and *Dataset2* reveals that most of the bot accounts are suspended and many accounts are not in English. In order to ensure genuine representation both from humans and bots accounts, we collected an additional dataset as follows.

*Human accounts:* We collected human user accounts who directly engage with the Twitter handles of organisations such as university and have correspondences in terms of replies to the users' queries. This is a useful technique to discount for bot accounts since bots may find it difficult to engage in meaningful conversations. To the best of our knowledge, this is the first study to employ this approach to ensure the genuineness of human users. For the *social bots accounts*, we collected 500 bot accounts using a publicly available bot detection system known as *Botometer*[6] (Davis et al., 2016). The *Botometer* returns a probable bot account which may result in many false positives. To mitigate that, we manually annotated the results of *Botometer*. The annotators scanned through the accounts and labelled account as bot based on the following criteria: (1) the account should be active, not suspended or deleted and posting tweets in English only (2) if the account's screen name appears to be auto-generated e.g. *2jo120, 2jo24 and 37Hkyjdtytyhjgh* (3) if the profile picture of the account shows no obvious relationship with the account's posts e.g. account tweeting on Brexit but with *storm-trooper* profile picture (4) number of URLs or hashtags: if the tweets mostly consist of URLs or hashtags exceeding 70% of the content (5) activity interval: posting at least 15 tweets per minute.

The annotation process is quite laborious which explain why the small size in *Dataset3*. We are not particularly interested in collecting a very high number of accounts but a high number of tweets from real human and bot accounts. We used the Twitter API to collect a maximum of 1000 tweets from each account.

Table 1. Summary of datasets utilised in the study. Dataset2 is categorised in groups based on the number of the followers in each group (*k* and *m* denote thousand and million respectively)

| Category | Bot Accounts | Human Accounts | Bot Tweets | Human Tweets |
| --- | --- | --- | --- | --- |
| Dataset1 | 685 | 641 | 27,766 | 5,341 |
| Dataset2 1k | 75 | 76 | 22,432 | 116,576 |
| Dataset2 100k | 266 | 343 | 112,387 | 98,271 |
| Dataset2 1m | 137 | 184 | 45,605 | 53,700 |
| Dataset2 10m | 25 | 25 | 9,062 | 11,869 |
| Dataset3 | 100 | 100 | 83,976 | 74,483 |

## 4.2 Classification of Account as Bot or Human Using Lexical Features

We use machine learning to measure the extent at which these features aid in identifying bot accounts. We train a number of classifiers, namely K-nearest neighbour (KNN), Naive Bayes, Support Vector machine for Classification (SVC) and Random Forest on the three datasets. As shown in Table 3, three experiments were conducted for *Dataset2*: (1) using our lexical features, denoted as *L,* (2) using the features in Table 2 from (Gilani et al., 2017), denoted as *F, and* (3) a combination of (1) and (2), denoted as *FL*. The proposed lexical features (L) in this study are *TTR, lexical diversity, average number contractions and emoticons*.

The classifiers were built and trained using *scikit-learn*[7] (Pedregosa et al., 2011), a machine learning toolkit supported by Google and *INRIA*[8]. Stratified 10-fold cross-validation was used to measure the overall

---
[6] botometer.iuni.iu.edu/#!/api
[7] scikit-learn.org/stable
[8] inria.fr/en

*accuracy, precision, recall and roc-auc score* of each classifier. The Random Forest classifier performs best as shown in Table 4.

Table 2. Features used in *Dataset2*

| Features |
| --- |
| Age of account, Favourites-to-tweets ratio, Lists per user, Followers-to-friends ratio, User favourites, Likes/favourites per tweet, Retweets per tweet, User replies, User tweets, User retweets, Tweet frequency, URLs count, Activity source type[S1= browser, S2= mobile apps, S3= OSN management, S4= automation, S5= marketing, S6=news content, S0= all other], Source count, CDN content size |

Table 3. Datasets and respective description of features utilised in training the prediction model

| Dataset | Description |
| --- | --- |
| Dataset2_F | features in Table 2 |
| Dataset2_FL | features from in Table 2 and our proposed lexical features (L) |
| Dataset2_L | proposed lexical features (L) |
| Dataset3_L | proposed lexical features on Dataset3 |

## 5. RESULTS AND DISCUSSION

The following section presents the main findings of the study. Figure 1 shows empirical evidence that the proposed lexical features are among the important features for the identification of automated accounts. Similarly, Figures 2, 3 and 4 show how lexical features manifest in humans and bot accounts.

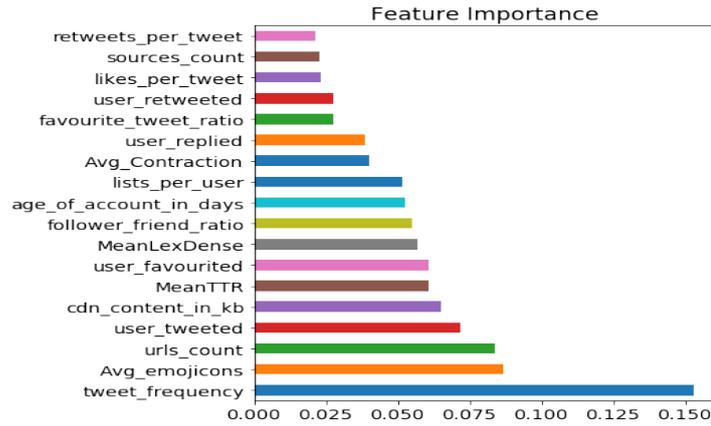

Figure 1. A comparison of importance of the proposed lexical features and features utilised in a related study

*Lexical Diversity:* Figure 2 shows the results of computing lexical diversity in human and bot accounts. *Lexical diversity* is expected to be higher in humans, since humans have been shown to generate better and novel content on Twitter (Gilani et al., 2017). However, this is not entirely true, especially in some automated accounts by prominent organisations, as shown in Figure 2. Accounts with a higher number of followers under the bot category are shown to have higher lexical diversities than the corresponding human counterparts. This is probably because such accounts are managed to update a large number of users on various topics on regular basis. Accounts in this category include organisational accounts, such as the BBC, politicians or popular celebrities. However, in *Dataset3*, humans have higher lexical densities which can be linked to the approach that was employed for the data collection. The dataset is a representative of an average user on Twitter. This confirms our earlier intuition that a typical human user account is expected to have a higher lexical diversity.

*Usage of Contractions:* Figure 3 shows how the usage of contraction varies across the datasets. With the exception of *Dataset3*, the difference in the use of contracted words between humans and bots is not very significant. This can be due to the fact that users with many followers on Twitter, such as organisational accounts or politicians, tend to use contracted words in tweets. With the exception of *Dataset3*, the difference in the use of contracted words between humans and bots is not very significant. This can be attributed to the fact that users with a high number of followership on Twitter will tend to use contracted words, e.g. (*Dataset2 1M and Dataset2 10M*), which mainly contain tweets of organisations, celebrities or politicians.

*Usage of Emoticons:* We found that the usage of *emoticons* is higher in bot accounts than in human accounts across all the different datasets as shown in Figure 4. Surprisingly, the results suggest that bot accounts utilise more *emoticons* in their tweets than humans. This is contrary to our prior intuition that humans are more likely to use more emoticons.

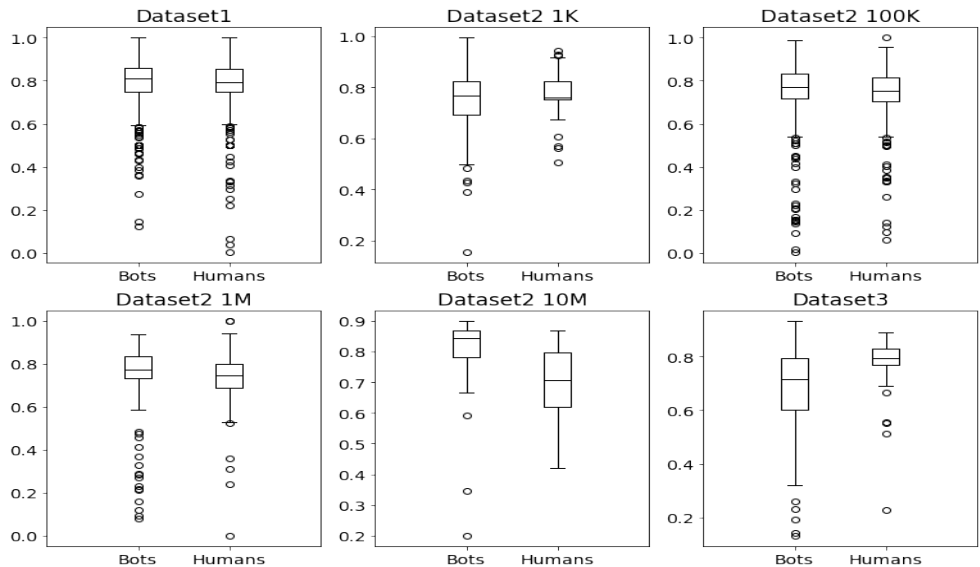

Figure 2. Average *lexical diversities* of *human* and *social bot* accounts

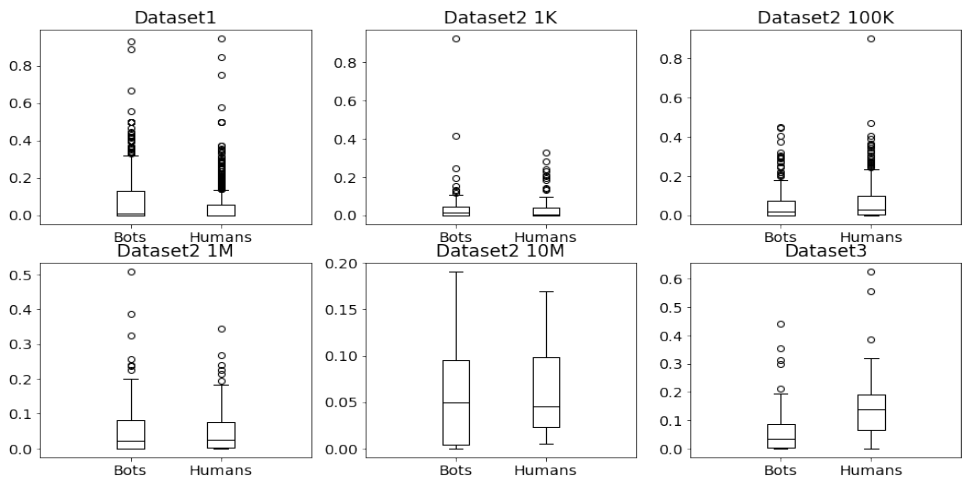

Figure 3. Average Contractions across datasets utilised in this study

*Classification of bot and human accounts using lexical features:* We use machine learning algorithms to measure the extent at which our proposed lexical features aid detection of bot accounts. Table 4 shows the results of a trained random forest classifier across the datasets. Using the lexical features only we achieve an accuracy of 86% and AUC score of 87% in Dataset_3. In Dataset2_FL we achieve an AUC score of 95% which is a significant improvement over 71% achieved using only the features utilised in (Gilani et al., 2017).

The primary goal is to improve detection of bot accounts by adding lexical features into the detection system. *Emoticons* happen to be the most distinctive features between humans and bots. In Figure 4 we observe an agreement in the usage of *emoticons* in all the datasets, i.e. bots use *emoticons* more frequently than humans. It is evident from the classification result (Table 4) that lexical features significantly improve the performance of the detection system. The distinguishing power of *lexical features* appears to be less effective in *Dataset1* and *Datatset2* in relation to *Dataset3*. With extra measures during data collection and annotation, this effect can be mitigated. This is evident from *Dataset3* which was manually inspected, and emphasis should be placed in the collection of more robust and representative datasets for effective detection. Despite the variations, which we attribute to many false positives in *Dataset1* and *Datatset2*, *lexical features* prove to be strong indicators as shown in Figure 1. The figure shows that *emoticons (*captioned as *avg_emojicons)* is the second most important feature among the 18 features. The lowest performing of our proposed lexical features (*avg_contraction*) outperforms 6 features utilised in a related study.

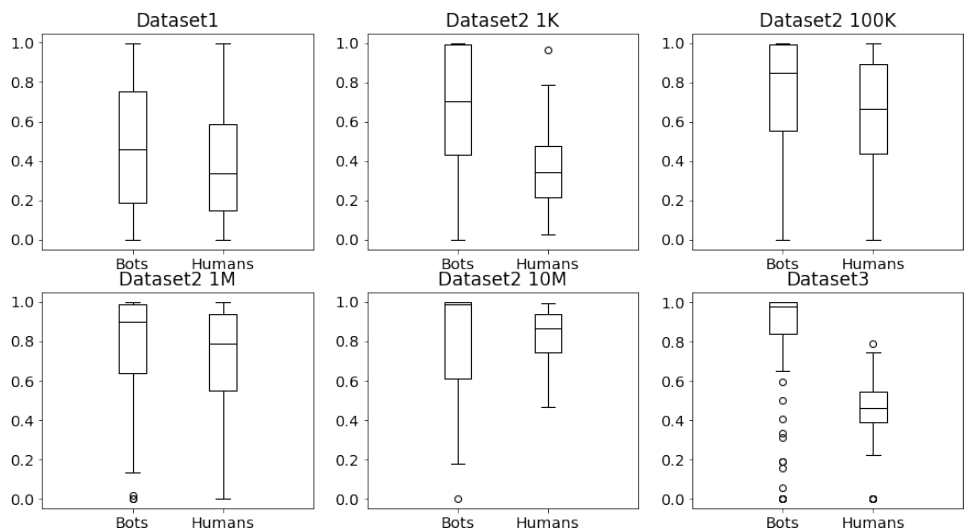

Figure 4. Average *emoticons* in the different datasets

Table 4. Datasets and respective prediction performances

| Dataset | *Accuracy* | *Precision* | *Recall* | *AUC Score* |
|---|---|---|---|---|
| Dataset1_L | 0.65 | 0.65 | 0.65 | 0.65 |
| Dateset2_F | 0.71 | 0.72 | 0.72 | 0.71 |
| Dateset2_L | 0.66 | 0.67 | 0.67 | 0.66 |
| Dataset2_FL | **0.95** | **0.96** | **0.96** | **0.95** |
| Dataset3_L | **0.86** | **0.87** | **0.87** | **0.87** |

## 6. CONCLUSION

Modern day social platforms have become part of our lives and effective social policing is required to ensure data credibility and civilised way of interaction. However, with the growing sophistication level of social bots, it is proving difficult to sanitise social platforms. The continuous increase in real-time streaming of tweets makes it practically ineffective to rely on many account features for detection. To effectively distinguish between a bots and a human user, an analysis of lexical richness of tweets posted by both users provides additional distinctive features. We train diverse classifiers to evaluate the role of lexical features in the detection of bot accounts. The newly proposed features significantly improve detection accuracy.

In this study, we have shown the difference between humans and bots in terms of lexical diversity, usage of contraction and emoticons, based on three different datasets. Lexical diversity and contractions vary across the different datasets. Contrary to our intuition, social bots accounts utilise a large number of emoticons in comparison to human accounts. We consider only English tweets as we do not focus on how lexical features are applied to other languages. Further investigation on this will improve the universality of our approach.

# ACKNOWLEDGEMENT

The authors wish to thank Prof. Reiko Heckel of University of Leicester for his positive feedback. The third author has participated in this research work as part of the CROSSMINER Project, which has received funding from the European Union's Horizon 2020 Research and Innovation Programme under grant agreement No. 732223.